\documentclass[reprint,aps,prb,reprintnumbers,amsmath,amssymb,superscriptaddress]{revtex4-2}

\usepackage{graphicx}
\usepackage{dcolumn}
\usepackage{bm}
\usepackage{upgreek}

\usepackage{epstopdf} 
\begin{document}

\title{Exchange-bias dependent diffusion rate of hydrogen discovered from evolution of hydrogen-induced noncollinear magnetic anisotropy in FePd thin films}

\author{Wei-Hsiang Wang}
\affiliation{Department of Physics, National Taiwan Normal University, Taipei 116, Taiwan}
\author{Yu-Song Cheng}
\affiliation{National Synchrotron Radiation Research Center, Hsinchu 300, Taiwan}
\author{Hwo-Shuenn Sheu}
\affiliation{National Synchrotron Radiation Research Center, Hsinchu 300, Taiwan}
\author{Wen-Chin Lin}
\email{Corresponding author. Email address: wclin@ntnu.edu.tw}
\affiliation{Department of Physics, National Taiwan Normal University, Taipei 116, Taiwan}
\author{Pei-hsun Jiang}
\email{Corresponding author. Email address: pjiang@ntnu.edu.tw}
\affiliation{Department of Physics, National Taiwan Normal University, Taipei 116, Taiwan}

\begin{abstract}
Hydrogenation-induced noncollinear magnetic anisotropy is observed from the evolution of the magnetic domains in FePd alloy thin films using magneto-optic Kerr effect (MOKE) microscopy. MOKE images reveal  complicated competitions between different magnetic anisotropies during hydrogen diffusion into the film. An intriguing enhancement of the hydrogen diffusion rate due to the presence of an initial exchange bias induced by a high magnet field is thereby discovered, pointing to an additional scope of controllability of magnetic metal hydrides as potential future hydrogen sensing and storage materials.
\end{abstract}
\maketitle

\section{Introduction}

The demand for stable hydrogen sensing and storage materials is growing because of increasing importance of fuel cell applications. Among the proposed hydrogen sensors, metal-hydride systems have been widely studied in the last decades, because hydrogenation of metals can modify the  morphologies, crystalline structures, or electronic structures, leading to modulated optical and magnetic properties \cite{Hjorvarsson2003,Remhof2008,Chang2020}. Therefore, hydrogenated metallic thin films are considered as potential functional materials in high-tech applications \cite{Remhof2008}. Pd, in particular, are often incorporated in metal hydrides as a catalyst to facilitate hydrogen absorption and desorption \cite{Remhof2008,Adams2011,Al-Mufachi2015,Silva2012}. Combinations of Pd layers and other functional thin films, such as Co/Pd based devices, are being investigated for their hydrogenation effects. As hydrogenation is found to be able to manipulate noncollinear magnetic states by reversibly tuning the Dzyaloshinskii--Moriya interaction \cite{Chen2021,Sandratskii2020,Yang2020,Hsu2018}, related research is drawing more attention for possible applications. The applicability of CoPd heterostructures or alloys as hydrogen sensors has recently been demonstrated through hydrogen-dependent variation of the magnetic anisotropy observed via ferromagnetic resonance \cite{Chang2013,Causer2019} or magneto-optic Kerr effect (MOKE) microscopy \cite{Chang2019}. 

The underlying mechanism of the magnetic anisotropy changing with the concentration and position of hydrogen, however, is not yet fully understood. In particular, hydrogen absorption has been demonstrated by many studies to lead to a reduction of perpendicular magnetic anisotropy (PMA) in CoPd alloy or multilayer films and nanostructures \cite{Causer2019,Munbodh2011,Chang2017,Chang2018}, and the reduction of PMA can be attributed to either magnetoelastic effects or modification of the electronic structures and orbital moments at the interface between Co and Pd \cite{Munbodh2012,Lueng2016}. However, a few other experiments on CoPd have revealed an enhancement of PMA instead \cite{Okamoto2002,Lin2016}. More research is required to completely understand the phenomenon.

The exchange bias (EB) can also be affected by hydrogenation. The study of EB has been of great interest because of its ability to pin the magnetization of a reference layer in a magnetic memory device. Setting an EB is often achieved by cooling a ferromagnetic (FM)/antiferromagnetic (AFM) bilayer under an applied field at the Néel temperature of the AFM \cite{Nogues1999}, and the EB typically manifests itself as a unidirectional field shift (EB field) of the hysteresis loop, which is a result of the interplay of the anisotropy, the  coupling between the two layers, and the external magnetic field \cite{Camarero2005,Chung2005,Nogues1996,Bruck2008,Blomqvist2005}. Since insertion of hydrogen atoms into the films is known to change the magnetic ordering, sizable exchange bias fields have then been observed in FM/AFM systems upon hydrogenation \cite{Wang2018}. The possibility of inducing EB has also been demonstrated recently in systems composed of two FM layers with one layer exhibiting in-plane anisotropy and the other exhibiting out-of-plane anisotropy, in which EB is observed in the layer with in-plane anisotropy without the necessity of a field cooling procedure \cite{Sort2004,Bollero2006,Bollero2008,Navas2012,Nguyen2011}. Further manipulation of the EB by utilizing hydrogenation is therefore expected to provide more insight in the FM/FM bilayer systems.

The kinematics of hydrogen diffusion in metals is crucial for achieving high stability and sensitivity for a hydrogen sensor. Hydrogen-induced changes in optical properties, especially in transmission and reflectivity, provide a noninvasive method of detecting hydrogen diffusion in materials such as VH$_x$, YH$_x$, and MgH$_x$ \cite{Huiberts1996,denBroeder1998,Kerssemakers2000,Remhof2002,Palsson2012,Karst2020}. However, investigation of the kinetics of hydrogen diffusion in most nontransparent metallic materials remains a challenge. Reversible changes of magnetic properties of ferromagnetic thin films such as magnetic Pd alloy films induced by hydrogenation are therefore being intensively investigated for the study of hydrogen diffusion. Among the measurement techniques of magnetic properties, MOKE microscopy has long been broadly applied in probing nanoscale structures because of its high sensitivity and feasibility in monitoring the evolutions of the magnetic domains, and therefore becomes a powerful tool to track hydrogen diffusion \cite{Chang2019}. 

Besides CoPd, FePd compounds have also attracted much attention in light of its high resistance to corrosion \cite{Kryder2008}, PMA due to the tetragonal \textit{L}1$_0$ phase \cite{Wei2009}, and the existence of Néel-type skyrmions in FePd monolayers \cite{Romming2013}. As hydrogenation of CoPd is being actively studied to explore the underlying mechanisms, FePd multilayer or alloy films should provide an additional scope in that the diffusion time of hydrogen into FePd before saturation is much longer than that into CoPd possibly because of the difference in their crystalline structures \cite{Chang2020,Lin2012,Lin2013}, without compromising the sensitivity of the magnetic properties to hydrogen absorption. This allows us to monitor the influence of hydrogen absorption on the magnetic properties as a function of time as hydrogen diffusion is taking place, making it possible to differentiate between various noncollinear anisotropies at different stages.

In this study, we report the observation of the effects of hydrogen diffusion into a FePd thin film on the magnetic properties through MOKE microscopy accompanied with simultaneously recorded hysteresis loops. The evolution of the magnetic domains under the change of magnetic anisotropy with the hydrogen diffusion time reveals an unexpected strong enhancement of the hydrogen diffusion rate if an EB is induced by a high initial magnetic field before hydrogen exposure. Our study also demonstrates clear competition between different magnetic anisotropies during hydrogen diffusion into the film. A different spontaneous EB induced by hydrogenation is discovered in the experiment as well. Our results extend the picture of hydrogen diffusivity to the regime of magnetic and spintronic properties and enrich the design of future hydrogen sensing and storage devices.

\section{Material and Methods}

The 30-nm-thick FePd (40:60) alloy films were deposited on Al$_2$O$_3$(0001) substrates by \textit{e}-beam heated coevaporation in an ultrahigh vacuum chamber with a base pressure of $3\times10^{-9}$ mbar. Two evaporation guns were both aligned at 45$^{\circ}$ to the normal, and lay on the plane perpendicular to the [1120] axis of the Al$_2$O$_3$(0001) substrate to optimize the hydrogenation effects in the later experiment \cite{Chang2022}. This oblique deposition geometry allows uniaxial magnetic anisotropy (UMA) to be developed on the surface plane \cite{Chi2012}. The alloy compositions and film thicknesses were controlled by the respective deposition rates of the elements, and were calibrated by Auger electron spectroscopy, x-ray photoelectron spectroscopy, atomic force microscopy, and transmission electron microscopy with energy dispersive x-ray spectroscopy \cite{Hsu2017}. The samples were stored in a vacuum desiccator (internal pressure $\sim$0.15 bar) to sufficiently slow down oxidation for extending the reliability of later measurements.

\section{Experimental Results and Discussion}
\subsection{Characterization of the hysteresis loops}

\begin{figure}
\centering
\includegraphics[width=0.4\textwidth]{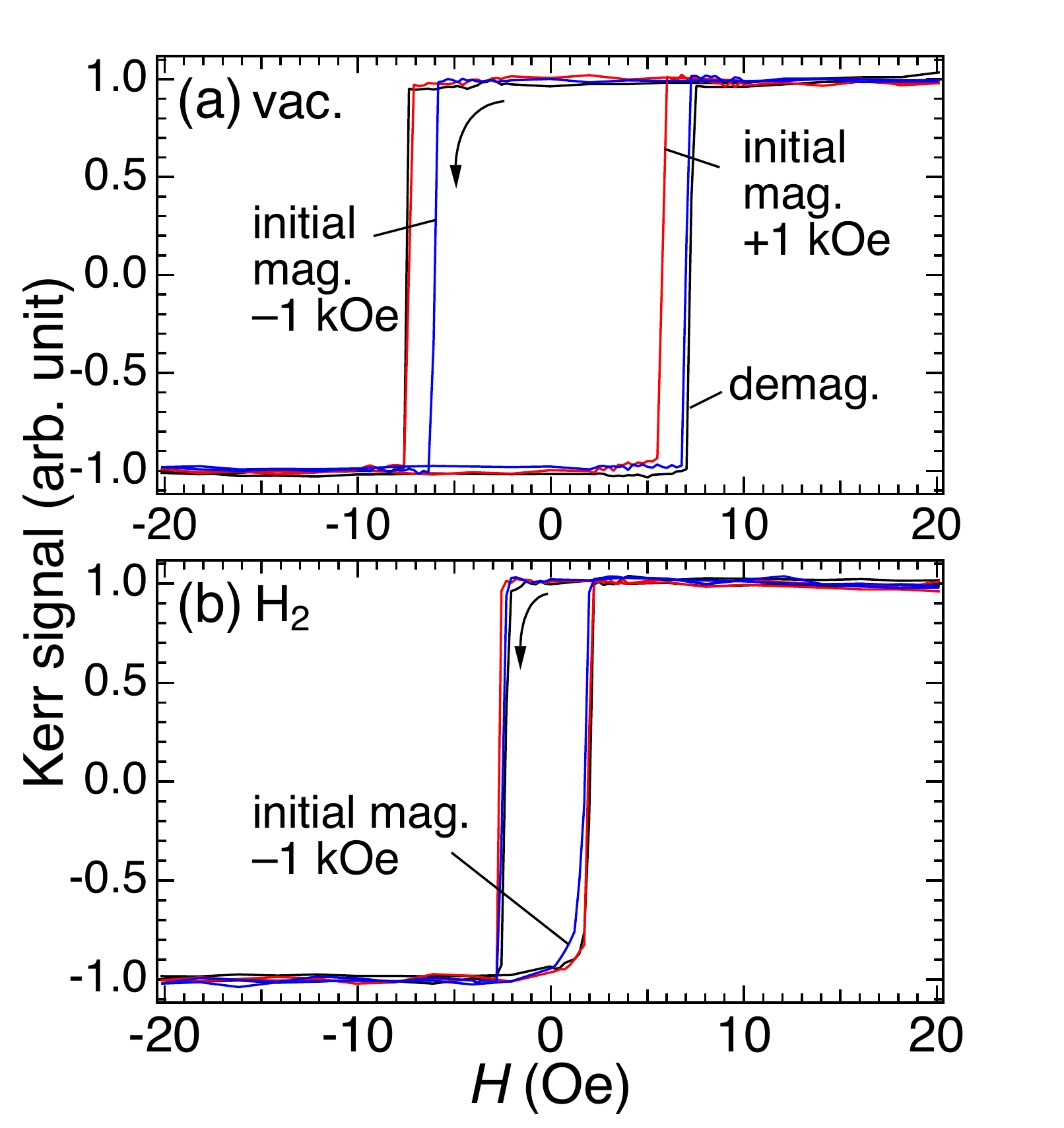}
\caption{Longitudinal MOKE hysteresis loops of the FePd film (a) in vacuum and (b) in hydrogen of 1.0 bar, respectively, with the Kerr sensitivity and the magnetic field applied along the magnetic easy axis. Each figure presents three loops measured with the sample previously magnetized under high field of $-1$ kOe and $+1$ kOe, and with the sample thoroughly demagnetized, respectively.}
\label{EB}
\end{figure}

The FePd thin film is investigated using longitudinal MOKE microscopy with the Kerr sensitivity along the magnetic field to capture the images of the magnetic domains during measurements of its hysteresis loops \cite{Chang2019}. The MOKE measurements are conducted on a region in middle of the film, but the evolution of the magnetic properties presented in the paper is general for all the regions on the entire film. Presented in Figs.~\ref{EB}(a) and \ref{EB}(b) are the MOKE hysteresis loops of the film measured in vacuum and in clean hydrogen soon after the pressure reaches 1.0 bar, respectively, with an in-plane magnetic field ($H$) applied along the magnetic easy axis of the film. The blue loops are obtained after initial magnetization at a high field of $-1$ kOe along the easy axis, the red ones are after initial magnetization at $+1$ kOe,  and the black ones are after demagnetization. Initial magnetization is performed before hydrogen exposure for Fig.~\ref{EB}(b). In Fig.~\ref{EB}(a), both blue and red loops have an average coercivity ($H_\mathrm{c}$) of 6.5 Oe, accompanied with an apparent horizontal shift indicating the existence of an EB in the system. The blue loop initially magnetized at $-1$ kOe has a positive EB field of $+0.5$ Oe, and the red one initially magnetized at $+1$ kOe has a negative EB field of $-0.8$ Oe, demonstrating a negative EB effect. The EB can be erased by applying a strong magnetic field of 8 kOe perpendicular to the film or an ac field of 3 kOe along the easy axis and then turning the field off, resulting in a ``demagnetized" hysteresis loop shown as the black curve in Fig.~\ref{EB}(a), with a larger $H_\mathrm{c}$ of 7.3 Oe. The EB of the FePd film can be attributed to the domain-pinning effect at the Pd-rich/FePd interface due to PMA in the top Pd-rich layer \cite{Wang2020}. The Pd-rich layer has been created because of a mild oxidation of Fe, which removes some Fe from the upper part of the film to form oxides on the surface and therefore increases the Pd concentration near the top. The PMA in the Pd-rich layer further stabilizes the in-plane domains in FePd, leading to a unidirectional anisotropy (UDA) and an EB \cite{Sort2004, Navas2012}.

After each loop measurement in Fig.~\ref{EB}(a), the chamber is filled with hydrogen gas to 1.0 bar in 30 sec, and a loop measurement is repeated in 10 sec after the pressure reaches 1.0 bar. Each loop takes 120 sec to finish \cite{dehydrogenation}. The resulting hysteresis loops are significantly modified when the sample is exposed to hydrogen, as shown in Fig.~\ref{EB}(b).  Each of the three loops shows a remarkable decrease in $H_\mathrm{c}$ to a value of $\sim$1.7 Oe, and the apparent EB in the two initially magnetized cases disappears. The disappearance of EB may be attributed to the reduction of PMA caused by hydrogenation \cite{Munbodh2011,Causer2019,Klyukin2020}, leading to a diminished UDA. Instead, a new spontaneous EB field of a very small value $\sim$$-0.05$ Oe is generated and observed in all three loops in Fig.~\ref{EB}(b), including the initially demagnetized one. The emergence of the spontaneous EB can be more clearly confirmed when reproduced at a lower hydrogen pressure with a lower diffusion rate (see Supplemental Material Sec.~I). These loops exhibit some asymmetry in that the reversal of the magnetization starting at the bottom-right corner of the square loop seems more rounded than the top-left corner, especially for the blue loop initially magnetized at $-1$ kOe, which originally has a positive EB field in vacuum. This hydrogenation-induced asymmetry indicates the existence of incomplete domains walls at a phase interface and a noncollinear anisotropy with unparalleled UMA and UDA \cite{Li2006,Jimenez2009}. Therefore, a small rotation of UMA is suspected to be induced in the system by hydrogenation, which will be further discussed in the following experiments and discussion.

\begin{figure}
\centering
\includegraphics[width=0.5\textwidth]{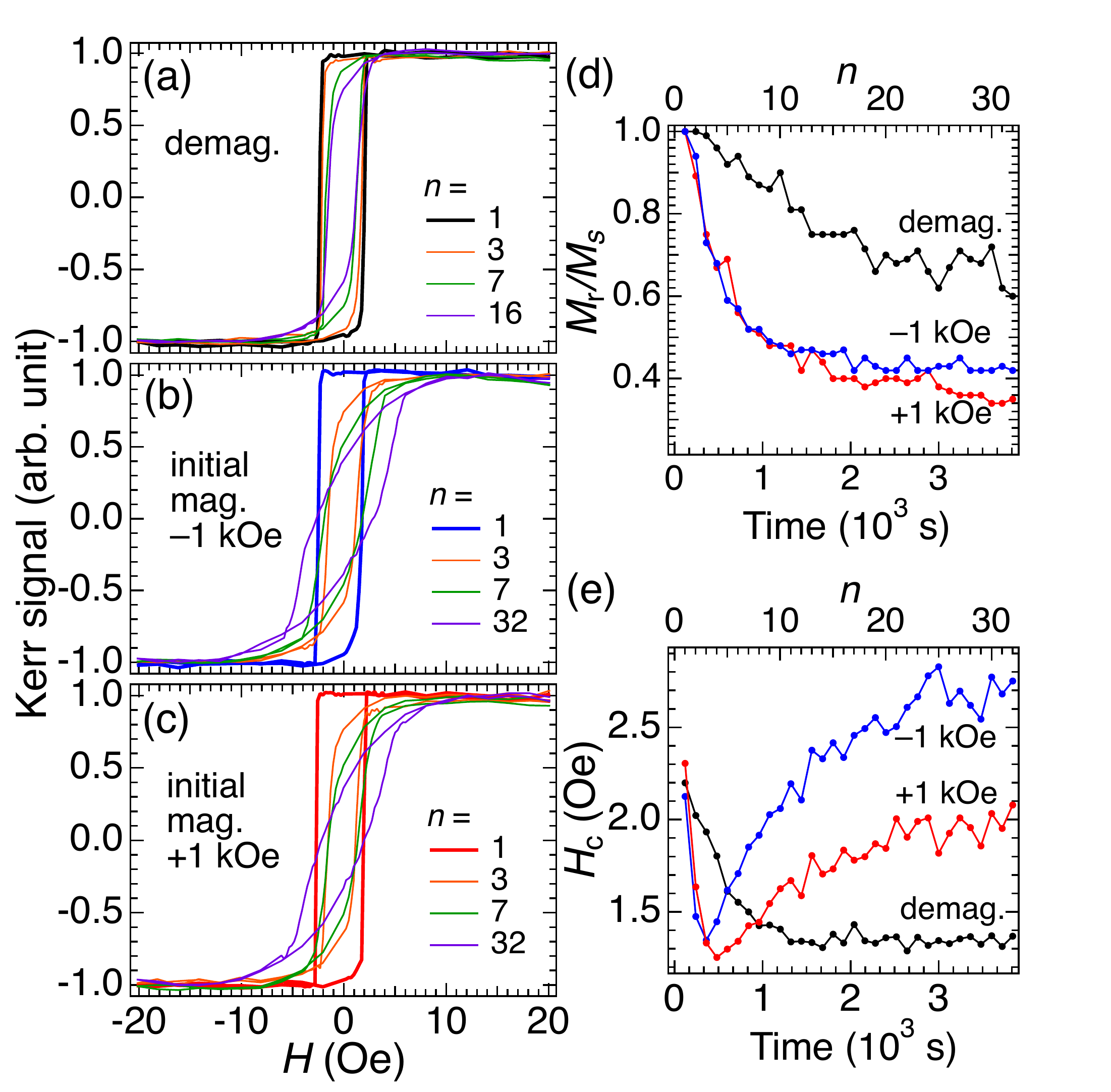}
\caption{Longitudinal MOKE hysteresis loops with the Kerr sensitivity and the magnetic field applied along the  original easy axis (i.e., the easy axis before hydrogenation) of the film. These loops are from $n=1$, 3, 7, and 16 in (a) the demagnetized case, and from $n=1$, 3, 7, and 32 in the cases initially magnetized at (b) $-1$ kOe and (c) $+1$ kOe, respectively. The dependence of $M_\mathrm{r}$/$M_\mathrm{s}$ and $H_\mathrm{c}$ on the hydrogen diffusion time are displayed in (d) and (e), respectively. The corresponding $n$ is labeled on the top axis. }
\label{sequence}
\end{figure}

Measurements of hysteresis loops in hydrogen are repeated 31 times after each loop in Fig.~\ref{EB}(b). The representative loops of the sequences for the demagnetized case and for the initially magnetized (under $\pm1$ kOe) cases are displayed in Figs.~\ref{sequence}(a)--\ref{sequence}(c), respectively. $n$ indicates the order in which the loop is measured. It can be seen that, for all three sequences, the shape of the loop changes from a square to a fusiform as $n$ increases, but is more tilted and plumped in the initially magnetized cases. The ratio of the magnetic remanence ($M_\mathrm{r}$) to saturation ($M_\mathrm{s}$) decreases as $n$ (and hence the diffusion time of hydrogen) increases, with its time evolution summarized in Fig.~\ref{sequence}(d). The decreasing rate of $M_\mathrm{r}$/$M_\mathrm{s}$ within the first 1000 sec in the initially magnetized cases is much higher than that in the demagnetized case, but notably slows down afterward as $M_\mathrm{r}$/$M_\mathrm{s}$ approaches a minimum of $\sim$0.4, whereas $M_\mathrm{r}$/$M_\mathrm{s}$ in the demagnetized case stays above 0.6 during the 3600-sec measurement. On the other hand, $H_\mathrm{c}$, as summarized in Fig.~\ref{sequence}(e), keeps decreasing gradually with time in the demagnetized case, until it reaches a minimum of $\sim$1.3 Oe in $\sim$1200 sec, and stays around the minimum afterward. However, in the initially magnetized cases, $H_\mathrm{c}$ decreases abruptly only within the first 240 or 360 sec to a minimum of $\sim$1.3 Oe, before it dramatically rises up afterward, with the increasing rate apparently higher in the case with initial magnetization under a negative high field (blue curve) than that under a positive high field (red curve). Another important feature found in all three sequences (Figs.~\ref{sequence}(a)--\ref{sequence}(c)) is that the asymmetry that emerges in the $n=1$ square loop gradually disappears as the diffusion time increases, when the loop shape becomes less square at larger $n$. Any of these loops can be reproduced even without repetitive field cycling. Therefore, the time evolution of the magnetic properties shown in Fig.~\ref{sequence} can be attributed to the effect of hydrogen diffusion instead of field training. The decrease in $H_\mathrm{c}$ upon hydrogenation can be attributed to the decrease in the anisotropy field $H_\mathrm{a}$ of the film \cite{Brown1945,Aharoni1962,Kronmuller1987,Kronmuller2003} (see Supplemental Material Sec.~II for measurements and detailed discussion on $H_\mathrm{a}$). The rebound of $H_\mathrm{c}$ from the minimum afterward to increase substantially with diffusion time in the initially magnetized cases, on the other hand, may be attributed to an increasing magnetocrystalline anisotropy (MCA), which is reflected by the increase of $H_\mathrm{a}$. The MCA will be further discussed in the following experiment. 

\subsection{Evolution of the magnetic domains}

\begin{figure*}
\centering
\includegraphics[width=0.76\textwidth]{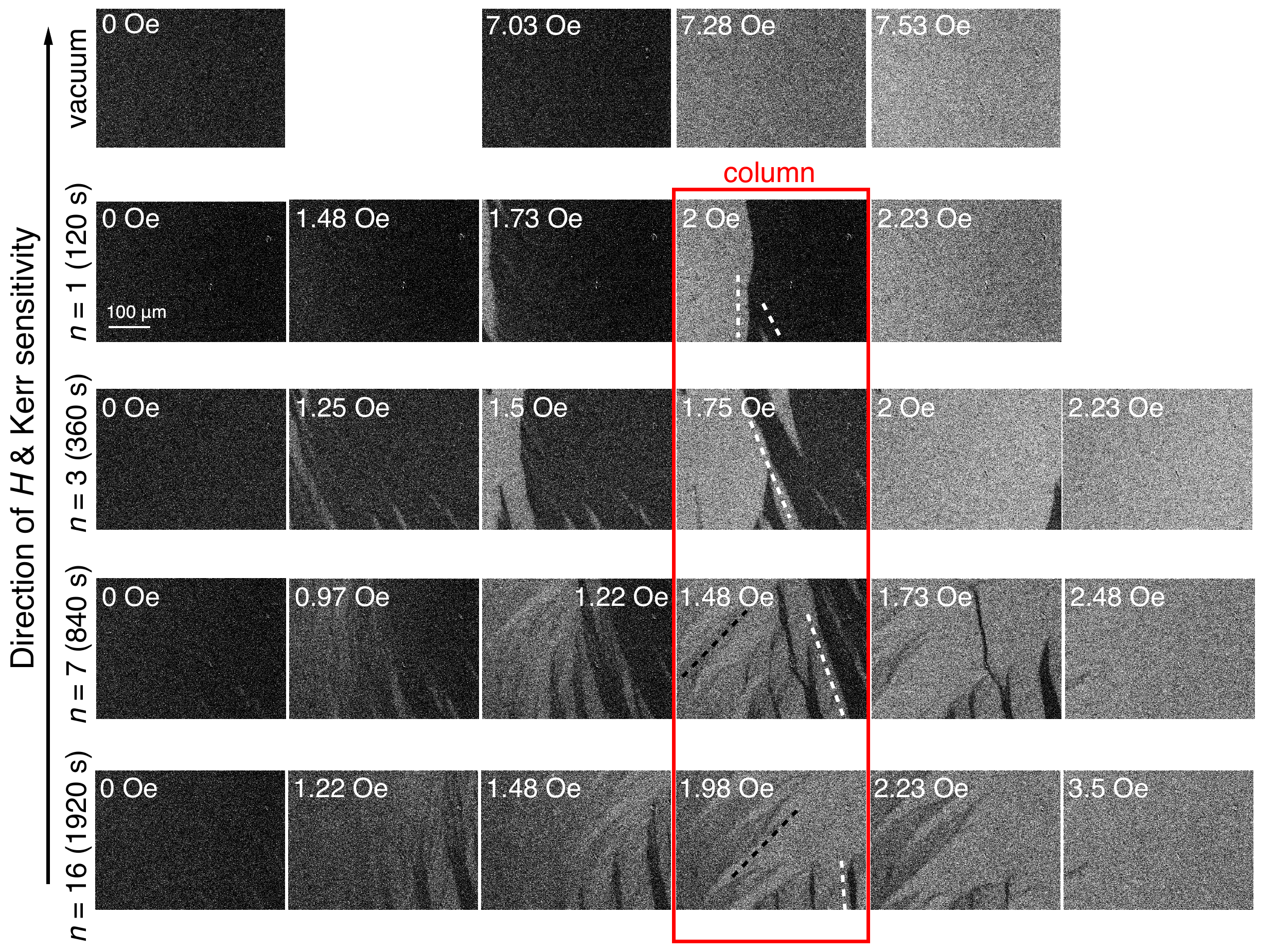}
\caption{Evolutions of the magnetic domain structures simultaneously recorded with the hysteresis loops in Fig.~\ref{sequence}(a) for the demagnetized case. Dashed lines are the eye-guiding lines for the orientations of the domain stripes representing different anisotropies.}
\label{sequenceDemag}
\end{figure*}

\begin{figure*}
\centering
\includegraphics[width=0.76\textwidth]{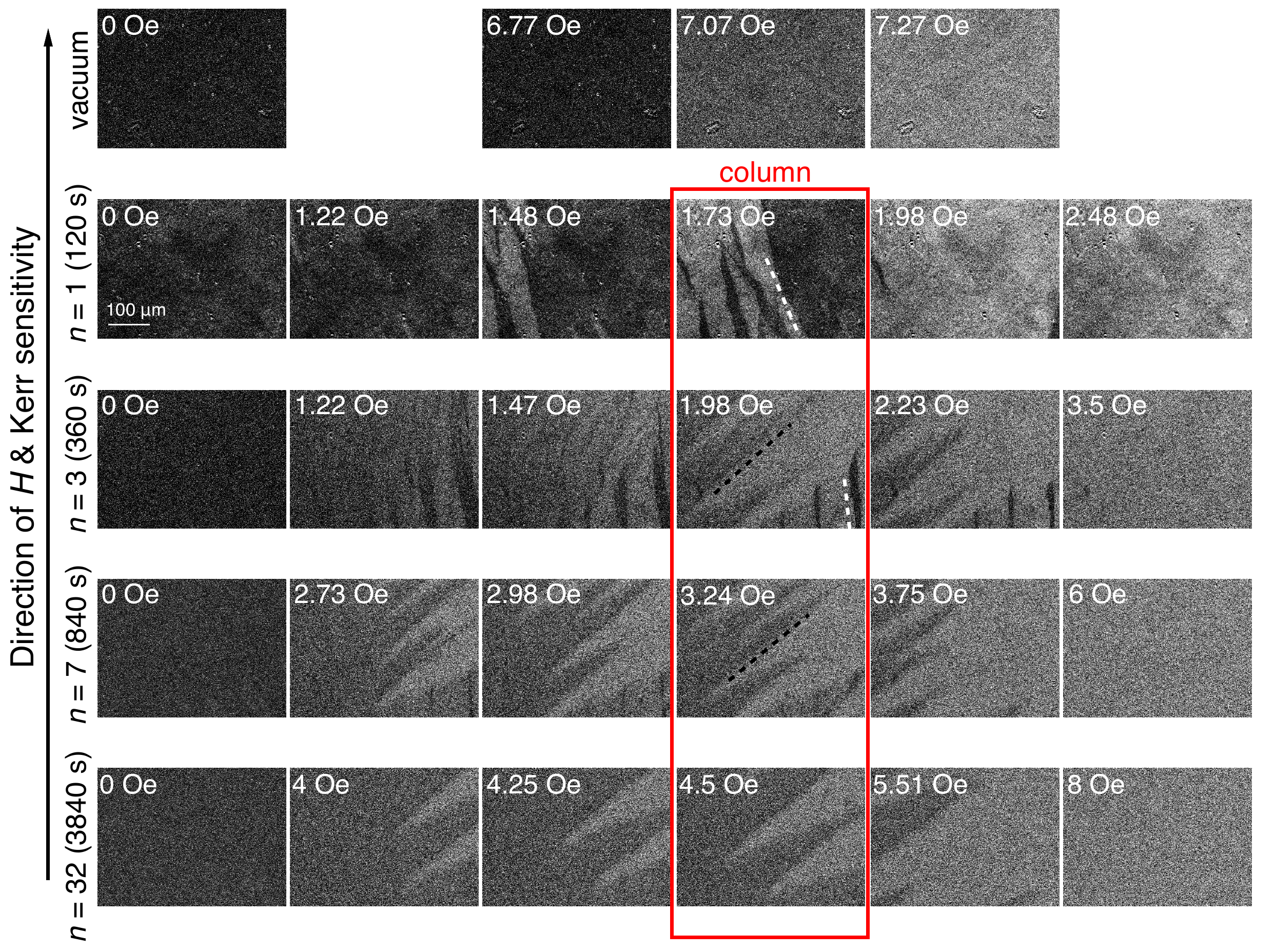}
\caption{Evolutions of the magnetic domain structures simultaneously recorded with the hysteresis loops in Fig.~\ref{sequence}(b) for the case with initial magnetization at $-1$ kOe. }
\label{sequenceHighn}
\end{figure*}

The effect of the hydrogen exposure can be more clearly seen from the evolution of the magnetic domains observed in the MOKE images. The magnetic domains simultaneously recorded with the hysteresis loops of the demagnetized case in Fig.~\ref{sequence}(a) and of the initially magnetized case in Fig.~\ref{sequence}(b) are displayed in Figs.~\ref{sequenceDemag} and \ref{sequenceHighn}, respectively, with increasing $H$ from left to right to accomplish magnetization reversal. When in vacuum, the magnetic moments of the entire film rotate coherently to reverse the whole magnetization without the occurrence of any observable magnetic-domain patterns, as shown in the first row of Figs.~\ref{sequenceDemag} and \ref{sequenceHighn}. After being exposed to hydrogen, however, complex domain patterns emerge during the magnetization reversal, and their dependence on the hydrogen diffusion time can be clearly tracked vertically through the images in the column boxed with red lines. Each image in the column is taken at a field when the total magnetization is rapidly changing, yielding complex domain patterns. The original easy axis in vacuum is along the vertical direction of the images. However, there is a hint of partial rotation of the easy axis upon hydrogen exposure with the appearance of a short segment of the light-gray domain at the bottom of $n=1$ image in Fig.~\ref{sequenceDemag}(column). The slightly rotated UMA becomes clearer in the domain patterns at $n=3$. The following $n=7$ series, on the other hand, features the emergence of another new anisotropy at $n=7$, where two sets of domain stripes with different orientations (indicated respectively with the white and black dashed eye-guiding lines) coexist. The two anisotropies are present in both $n=7$ and 16, exhibiting the competition between multiple magnetic structures in the sample. The noncollinear multidomain configuration is well known for single-layer films with high magnetic anisotropy dispersion. The time evolution is best interpreted with the penetration depth and the concentration of the hydrogen diffused into the film, both of which increase with time. The diffusion rate depends on the concentration gradient of hydrogen, and the vacancy and interstitial diffusion paths in the lattice. At $n=3$, only the top of the film is majorly hydrogenated, giving a domain pattern of dark stripes (indicated with the white dashed line). As time passes (with increasing $n$), the rest of the film starts to receive more hydrogen. The overall increase of the hydrogen concentration leads to modification of the magnetic properties of the whole film, resulting in the medium-gray stripes (indicated with the black dashed line at $n = 7$ and 16). The anisotropy that determines the orientation of the medium-gray domain stripes may depend on the crystalline structure, which can be affected by the substrate beneath the film via interfacial interaction. It is found in another experiment that the medium-gray stripes along with other hydrogenation effects are absent if the $\alpha$-Al$_2$O$_3$(0001) substrate is rotated by $90^{\circ}$ for the process of film deposition \cite{Chang2022} or is replaced by glass, which is non-crystalline. The crystalline structure of the substrate can affect the development of the deposited film lattice. If the [1120] axis of the Al$_2$O$_3$(0001) substrate lies on the plane of oblique deposition of Fe and Pd, or if a non-crystalline substrate is used that does not intervene a quality development of the crystalline axis of the film, the [1120] zone axis will be well developed in the FePd(111) alloy film especially for thinner film growth, leading to a magnetic anisotropy that is too stable to be altered via later hydrogenation. The hydrogenation effect on the magnetic properties can be clearly observed only if the [1120] axis of the Al$_2$O$_3$(0001) substrate does not lie on the plane of oblique deposition.

If the sample has been initially magnetized under a negative high field, the diffusion rate of hydrogen into the sample seems to  speed up drastically according to the MOKE images displayed in Fig.~\ref{sequenceHighn}(column). For $n = 1$ in Fig.~\ref{sequenceHighn}(column),  the slightly rotated easy axis already dominates the domain pattern. At $n=3$, the domain pattern well resembles that at $n=16$ at 1.98 Oe in the demagnetized case in Fig.~\ref{sequenceDemag}(column), exhibiting the same two anisotropies. This implies that the presence of an initial EB mainly affects the diffusion rate of hydrogen, but not the kinds of anisotropies that can be induced by hydrogenation. Looking back at Fig.~\ref{sequence}(e), the diffusion rate can also be inferred from the dependence of $H_\mathrm{c}$ on the diffusion time, with $H_\mathrm{c}$ showing an abrupt decrease upon hydrogen exposure in the initially magnetized cases, but only a slow decrease in the demagnetized case. For $n>3$, a rise of $H_\mathrm{c}$ is observed in the cases with initial EB, but not in the demagnetized case. By comparing the increasing rate of $H_\mathrm{c}$, one can further infer a higher diffusion rate in the case with a positive initial EB field in vacuum (due to initial magnetization at $-1$ kOe) than that with a negative initial EB field. At $n=7$ in Fig.~\ref{sequenceHighn}(column), the dark domain stripes along the slightly rotated easy axis almost disappear, leaving only the medium-gray domains with the domain wall now developed into a zigzag shape, which corresponds to a new orientation of the magnetic axes that is very different from the original one (see Supplemental Material Sec.~III for the angular dependence of the coercivity and the remanence before and after hydrogenation). The reorientation of the system anisotropy starts to get stabilized around $n=7$, as also implied in Fig.~\ref{sequence}(d) with $M_\mathrm{r}$/$M_\mathrm{s}$ approaching a minimum around 1000 sec. At $n=32$, the zigzag wall line has fewer turning points when the sample is almost saturated with hydrogen. The orientation of the zigzag domain wall is very sensitive to the direction of the easy axis of a system \cite{Pokhil1997,Cerruti2007}. The development of the medium-gray zigzag domain hence indicates an emergence of a new easy axis of the film. The hysteresis loops and domain patterns obtained from the film with an initial EB are robust against a regular demagnetization procedure alone. The effect of the initial EB on the magnetic properties can only be erased by demagnetization after a dehydrogenation process \cite{dehydrogenation} is executed. 

The two key mechanisms that can change the magnetic properties of the alloy film upon hydrogen absorption are electron transfer and magnetoelastic effects \cite{Weidinger1996}. \textit{In situ} x-ray diffraction (XRD) has been performed on the film to examine the crystalline structure, and no measurable expansion of the FePd lattice is observed after hydrogen absorption (see Supplemental Material Sec.~IV). This implies that magnetostrictive effects are very small in our samples. Therefore, the changes in the magnetic anisotropies must be a result of electron transfer between Fe and Pd upon hydrogen absorption \cite{Munbodh2011}. Electron transfers upon hydrogenation have also been identified in our previous studies to cause variations in magnetic properties in CoPd films \cite{Liang2017,Liang2018}.

\subsection{Discussion}

Our experiments demonstrate complicated competitions between the UMA caused by oblique deposition of the film,  the original UDA due to PMA in the Pd-rich layer that leads to EB observed in Fig.~\ref{EB}(a), the new UDA that corresponds the small spontaneous EB ($-0.05$ Oe) in Fig.~\ref{EB}(b), and a probable MCA from the crystalline nature, under the influence of hydrogenation. The alteration of the anisotropy energies can be attributed to the changes in the electronic structure of the ferromagnetic group-VIII elements due to electron transfer from Pd through orbital hybridization upon the formation of hydrides, which diminishes the contribution of the shape anisotropy (i.e., the UMA in our sample due to oblique deposition \cite{Bubendorff2006,Lisfi2002}), and promotes the in-plane MCA \cite{Klyukin2020,Liang2018,Chang2018}. The sample in vacuum originally possesses UMA and the original UDA. Hydrogenation then removes PMA and hence the original UDA as mentioned earlier in Fig.~\ref{EB}(b),  causes a small rotation of UMA as shown at smaller $n$ in Figs.~\ref{sequenceDemag}(column) and \ref{sequenceHighn}(column), and possibly modifies the spin phase that results in the new UDA. The new UDA may be generated at the interface between the hydrogenated top layer with hydrogenation-induced phase inhomogeneity and the bottom portion that is not hydrogenated yet, since a phase separation at an interface is known to result in a spontaneous EB \cite{Tang2006}. The slightly rotated UMA can be regarded as a first sign of the diminished UMA caused by electron transfer. One of the most interesting discoveries in this study is the significant increase of the hydrogen diffusion rate after initial magnetization under a high field before hydrogen exposure. This can be observed by comparing Figs.~\ref{sequenceDemag}(column) and \ref{sequenceHighn}(column) and reviewing Figs.~\ref{sequence}(d) and \ref{sequence}(e). As hydrogen atoms would most probably advance on minimum-energy paths for diffusion \cite{Chohan2018}, the dependence of the diffusion rate on the initial EB may indicate changes of minimum-energy paths due to variations of magnetic ordering. Then, at $n=7$ in Fig.~\ref{sequenceDemag}(column) and $n=3$ in Fig.~\ref{sequenceHighn}(column), MCA of the film emerges with the help of increased hydrogen concentration diffused into the sample, promoting the magnetocrystalline nature that was initially suppressed and unobservable because of the existence of the strong UMA before hydrogenation. In the mean time, the phase interface created upon hydrogen exposure becomes blurred with increasing diffusion time and decreasing concentration gradient of hydrogen, which leads to the disappearance of the new UDA and thus the disappearance of the asymmetry in the hysteresis loops at larger $n$ in Figs.~\ref{sequence}(a)--\ref{sequence}(c). At $n=7$ and 32 in Fig.~\ref{sequenceHighn}(column), the newly appearing MCA dominates the whole film as UMA lessens more, manifesting itself with the zigzag domain taking all over. The stronger MCA in the initially magnetized cases results in the rebound and enhancement of $H_\mathrm{c}$ after a longer diffusion time.

The strong dependence of hydrogen diffusion on the initial EB is very intriguing, as it allows additional controllability of the hydrogenation effect on a potential hydrogen sensing or storage material. The initial EB not only enhances the hydrogen diffusion rate significantly, but also lifts the original limit on the saturation state. Unlike the demagnetized case in which $H_\mathrm{c}$ stays at its minimum after saturation, the initially magnetized cases advance further into a state in which the fusiform-shaped loop becomes more tilted and plumped as shown in Figs.~\ref{sequence}(b) and \ref{sequence}(c), and $H_\mathrm{c}$ rebounds from the minimum and increases substantially as shown in Fig.~\ref{sequence}(e), eventually reaching a stage where only the zigzag domain dominates the FePd film [i.e., $n=7$ and 32 in Fig.~\ref{sequenceHighn}(column)]. It is not clear yet how the initial magnetic ordering affects hydrogen diffusion. It is known that an external uniform magnetic field has no influence on the diffusion of hydrogen in Pd \cite{Verbruggen1985}. Therefore, the dependence of hydrogen diffusion on the initial EB may be attributed to variation of the micromagnetic or spintronic structures inside the FePd film. More research is necessary to fully understand the underlying mechanism.

\section{Conclusions}

In summary,  we report discovery of the dependence of hydrogen diffusion rate on a magnetic property of a material. The hydrogen diffusion rate in FePd alloy thin films is significantly enhanced if a UDA is induced by a high initial magnetic field before hydrogen exposure. This unexpected phenomenon is discovered by monitoring the competitions between multiple magnetic anisotropies in an FePd thin film during hydrogen diffusion into the film. MOKE microscopy shows hysteresis loops and complex magnetic domains that are functions of the diffusion time of hydrogen. The appearance and later disappearance of the new UDA with an spontaneous EB, the slightly rotated UMA during an earlier stage of hydrogenation, and the MCA with zigzag domains that later emerges and eventually dominates the whole film can be attributed to variation of the magnetic properties caused by electron transfer induced by hydrogenation, as XRD shows no measurable expansion of the film lattice. The state of hydrogen diffusion is clearly visualized in our experiment, revealing the apparent difference in the diffusion rate between the samples without and with an initial EB. This discovery allows additional controllability of the hydrogenation effect on the magnetic properties of a magnetic metal-hydride. Our study provides valuable information for the application of magnetic metal hydrides in development of hydrogen sensing and storage devices.

\section*{Acknowledgements}

This study was sponsored by the Ministry of Science and Technology of Taiwan under Grants No.~MOST 107-2112-M-003-004, No.~MOST 108-2112-M-003-011-MY2, No.~MOST 109-2112-M-003-009, and No.~MOST 110-2112-M-003-019.


\begin{thebibliography}{65}
\bibitem [{deh()}]{dehydrogenation}%
  \BibitemOpen
  \href@noop {} {}\bibinfo {note} {Note that a ``dehydrogenation" process is
  performed on the film before any following loop measurement in vacuum
  displayed in Fig.~\ref{EB}(a), by putting the sample in another vacuum
  chamber with continuous pumping for a day to ensure thorough removal of
  hydrogen from the film.}\BibitemShut {Stop}%
\end{thebibliography}
\end{document}


\title{Exchange-bias dependent diffusion rate of hydrogen discovered from evolution of hydrogen-induced noncollinear magnetic anisotropy in FePd thin films\\ \, \\SUPPLEMENTAL MATERIAL}

\author{Wei-Hsiang Wang}
\affiliation{Department of Physics, National Taiwan Normal University, Taipei 116, Taiwan}
\author{Yu-Song Cheng}
\affiliation{National Synchrotron Radiation Research Center, Hsinchu 300, Taiwan}
\author{Hwo-Shuenn Sheu}
\affiliation{National Synchrotron Radiation Research Center, Hsinchu 300, Taiwan}
\author{Wen-Chin Lin}
\email{Corresponding author. Email address: wclin@ntnu.edu.tw}
\affiliation{Department of Physics, National Taiwan Normal University, Taipei 116, Taiwan}
\author{Pei-hsun Jiang}
\email{Corresponding author. Email address: pjiang@ntnu.edu.tw}
\affiliation{Department of Physics, National Taiwan Normal University, Taipei 116, Taiwan}

\maketitle

\section{Hydrogen-Induced Spontaneous Exchange Bias}

\begin{figure}
\centering
\includegraphics[width=0.5\textwidth]{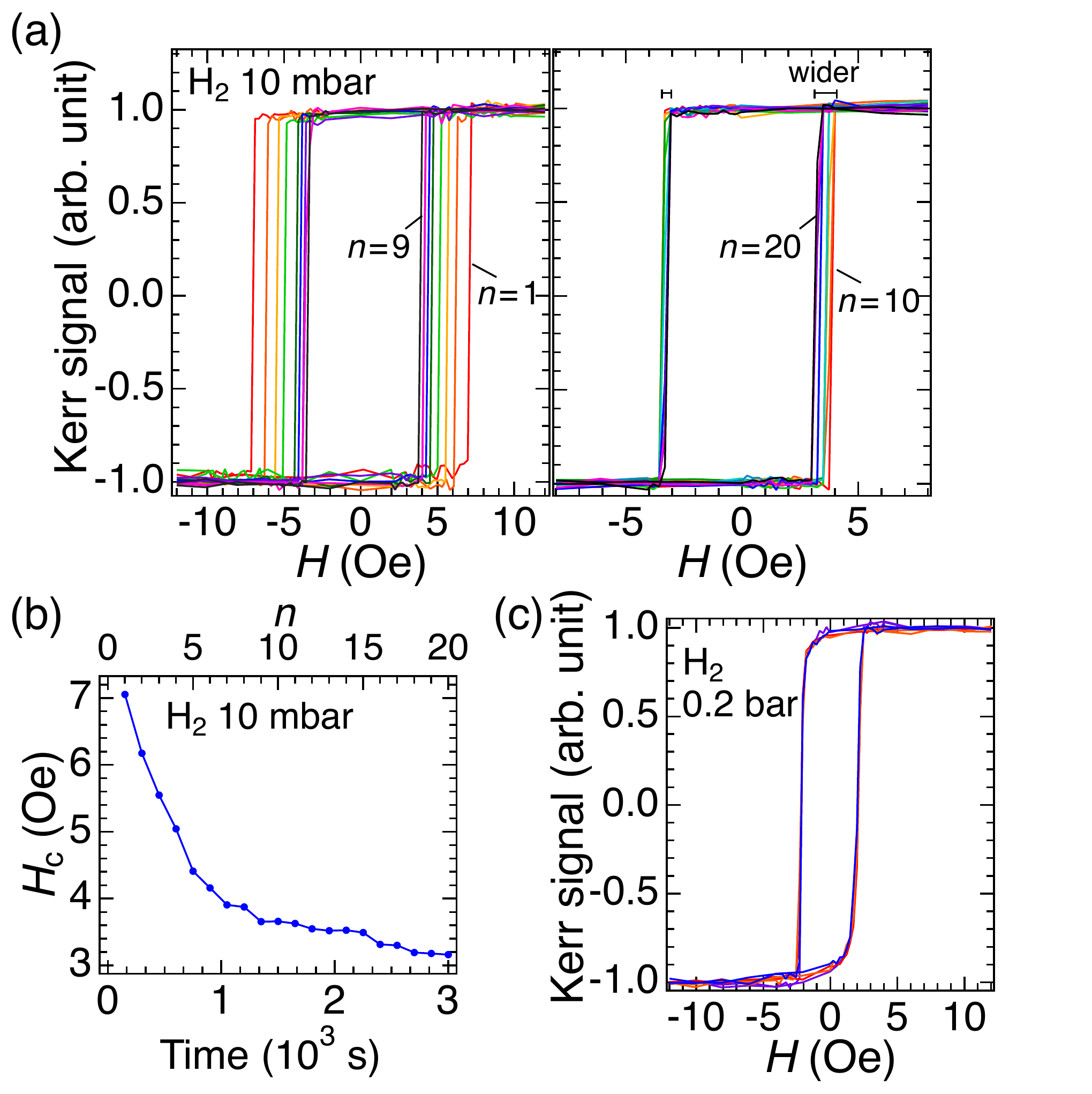}
\caption{(a) Longitudinal MOKE hysteresis loops under 10 mbar of hydrogen with the Kerr sensitivity and the magnetic field applied along the easy axis of a demagnetized FePd film. The dependence of $H_\mathrm{c}$ on the diffusion time is shown in (b). (c) Successive hysteresis loops under 0.2 bar of hydrogen with the magnetic field applied along the easy axis of the demagnetized sample.}
\label{lowH}
\end{figure}

The effect of hydrogen exposure on a demagnetized FePd film is also studied at lower hydrogen pressures and hence lower diffusion rates and hydrogen concentrations in the sample in order to confirm the mechanisms. Fig.~\ref{lowH}(a) shows successive hysteresis loop measurements under 10 mbar of hydrogen with $n=1$ to 9 on the left panel and $n=10$ to 20 on the right, each loop taking 150 secs to finish. The average $H_\mathrm{c}$ is found to be decreasing with time, with the relation shown in Fig.~\ref{lowH}(b). The decrease of the positive-field $H_\mathrm{c}$ becomes slightly faster than that of the negative-field one for $n>10$, as shown in the right panel, indicating that hydrogen absorption for $n>10$ induces some UDA in the sample. The appearance of this new UDA is consistent with our observation in Fig.~1(b) using 1.0 bar of hydrogen, where a small negative spontaneous EB of $-0.05$ Oe is present even in the demagnetized case. It is therefore speculated that hydrogenation washes out the EB induced by high-field initial magnetization shown in Fig.~1(a), and generates a new UDA with a small spontaneous EB in the film. The new UDA and the slight rotation of the easy axis shown at $n=1$ and 3 in Fig.~3, both induced by hydrogenation, lead to the asymmetry of the square loop shown in Fig.~1(b). This asymmetry is revisited in Fig.~\ref{lowH}(c), where the loops are measured at 0.2 bar instead right after the sample is saturated with hydrogen. These loops are found to stay asymmetric, indicating that the hydrogen saturation concentration at this pressure is high enough to induce asymmetry but low enough to avoid its disappearance, as opposed to the measurements in Fig.~2 at 1.0 bar, where the appearance and later disappearance of the loop asymmetry are both observed in the sequences. The shape of the asymmetric loops is persistent during several field cycles from the start to the end, indicating that the loop asymmetry is achieved by hydrogen exposure, not by field training. 

\section{Anisotropy Field}

\begin{figure}
\centering
\includegraphics[width=0.5\textwidth]{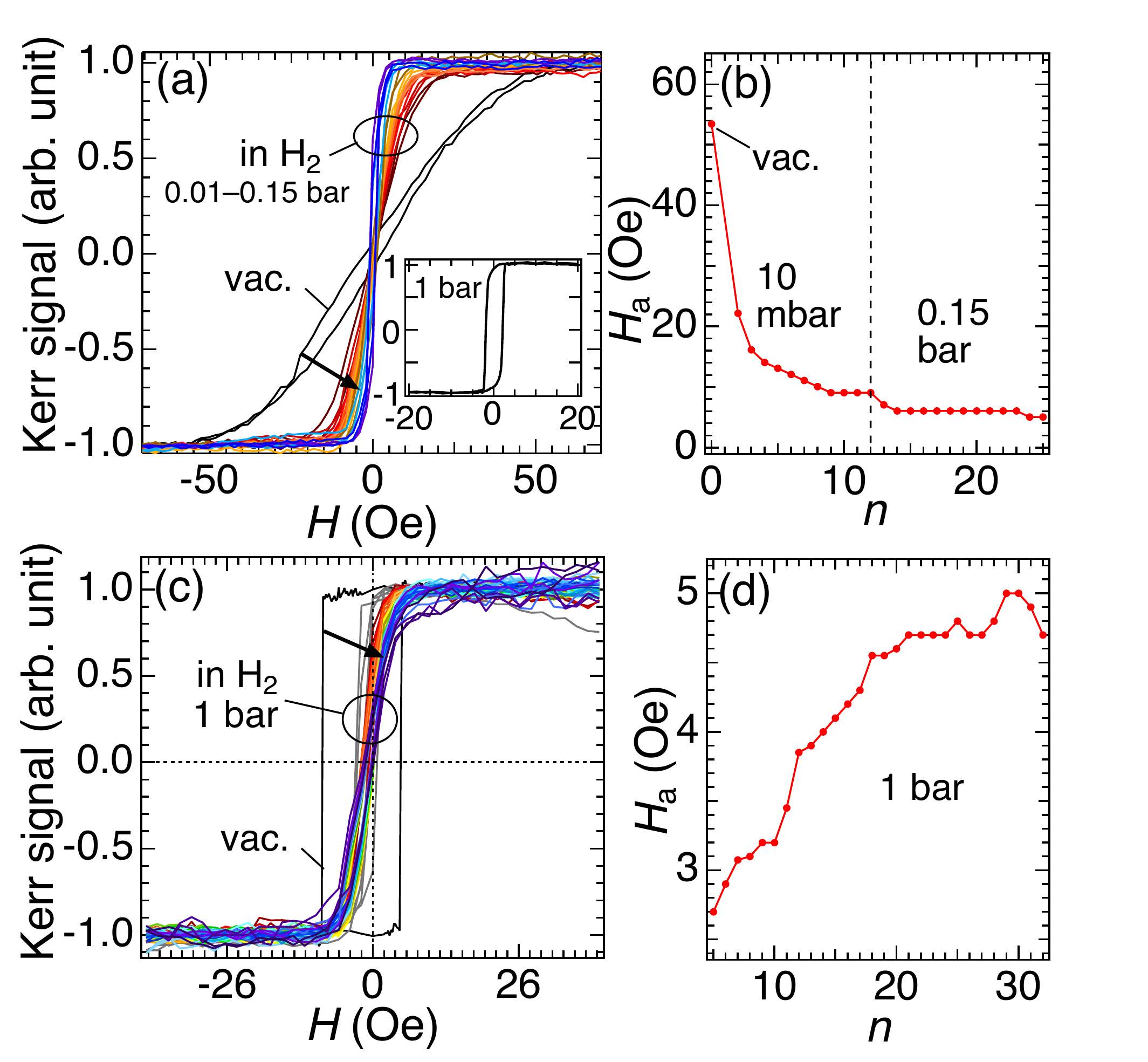}
\caption{(a) Successive longitudinal MOKE hysteresis loops of a demagnetized FePd film with the Kerr sensitivity and the magnetic field applied along the original, unrotated \textit{hard} axis. The first loop ($n=0$) is measured in vacuum. The following 12 loops are measured under 10 mbar of hydrogen (only $n=2$ to 7 and $n=12$ are shown) and then another 12 loops under 0.15 bar (only $n=13$, 17, and 24 are shown). Inset: longitudinal MOKE loop at 1 bar of hydrogen with the Kerr sensitivity and the magnetic field along the direction of the original hard axis. It is obvious from the square loop shape that the hard axis has deviated from its original orientation. (b) The dependence of $H_\mathrm{a}$ on $n$ from (a). (c) Successive longitudinal MOKE hysteresis loops measured respectively in vacuum and under 1-bar hydrogen ($n=1$ to 32) from a film that was initially magnetized at $+1$ kOe, with the Kerr sensitivity and the magnetic field applied along the \textit{new} hard axis. (d) The dependence of $H_\mathrm{a}$ on $n$ from (c) for $n \geq 5$.}
\label{Ha}
\end{figure}

Hysteresis loops of a demagnetized sample with the external magnetic field applied along the hard axis are also investigated, as shown in Fig.~\ref{Ha}(a). The first loop is measured in vacuum ($n=0$), then 12 successive loops under 10 mbar of hydrogen ($n=1$ to 12, with only $n=2$ to 7 and $n=12$ loops displayed), and then another 12 successive loops under 0.15 bar ($n=13$ to 24, with only $n=13$, 17, and 24 loops displayed), with their order and evolving trend indicated with an arrow in the figure. Each loop takes 240 secs. The corresponding average anisotropy field $H_\mathrm{a}$ estimated from the saturation field of each loop as a function of $n$ is displayed in Fig.~\ref{Ha}(b). It can be seen that hydrogenation causes a significant decrease in $H_\mathrm{a}$. When the decrease slowly stops as the sample is saturated with 10-mbar hydrogen, a raise in hydrogen pressure to 0.15 bar further decreases $H_\mathrm{a}$ in following measurements. A further raise in hydrogen pressure to 1 bar causes a rotation of the hard axis, as implied by the square shape of the 1-bar loop in the inset. It has been found theoretically and experimentally that $H_\mathrm{c}$ along the easy axis is closely related to $H_\mathrm{a}$, and is smaller than $H_\mathrm{a}$ for regular samples due to defects \cite{Brown1945,Aharoni1962,Kronmuller1987,Kronmuller2003}. Therefore, the decrease in $H_\mathrm{c}$ under hydrogenation observed in the demagnetized case and in the first $\sim$300 secs of the initially magnetized cases can be attributed to the decrease of $H_\mathrm{a}$, which reflects a decrease of UMA. 

Fig.~\ref{Ha}(c), on the other hand, shows the successive loops measured respectively in vacuum and under 1-bar hydrogen ($n=1$ to 32) for a film that was initially magnetized at $+1$ kOe, with the Kerr sensitivity and the magnetic field applied along the \textit{new} hard axis that is determined at $n=32$ when the sample is saturated with hydrogen. The significant difference in the loop shape before and after hydrogenation again indicates a reorientation of the magnetic axes. The $H_\mathrm{a}$ for $n \geq 5$ are summarized in Fig.~\ref{Ha}(d), and increases with $n$ until it approaches saturation. The increase in $H_\mathrm{a}$ reflects the increase in the dominating anisotropy MCA, which explains the increase of $H_\mathrm{c}$ in the initially magnetized cases shown in Fig.~2(e) after $n=3$.  

Fig.~\ref{Ha} as a whole demonstrates a decrease of UMA in an earlier stage of hydrogenation, followed by a substantial increase of MCA after $n\sim3$ under 1 bar. It is worth noting that the reorientation of UMA in a system with EB takes place only when UMA is small to allow the existence of incomplete domain walls in the system \cite{Jimenez2009,Morales2006}.

\section{Angular Dependence of the Coercivity and the Remanence}

\begin{figure}
\centering
\includegraphics[width=0.52\textwidth]{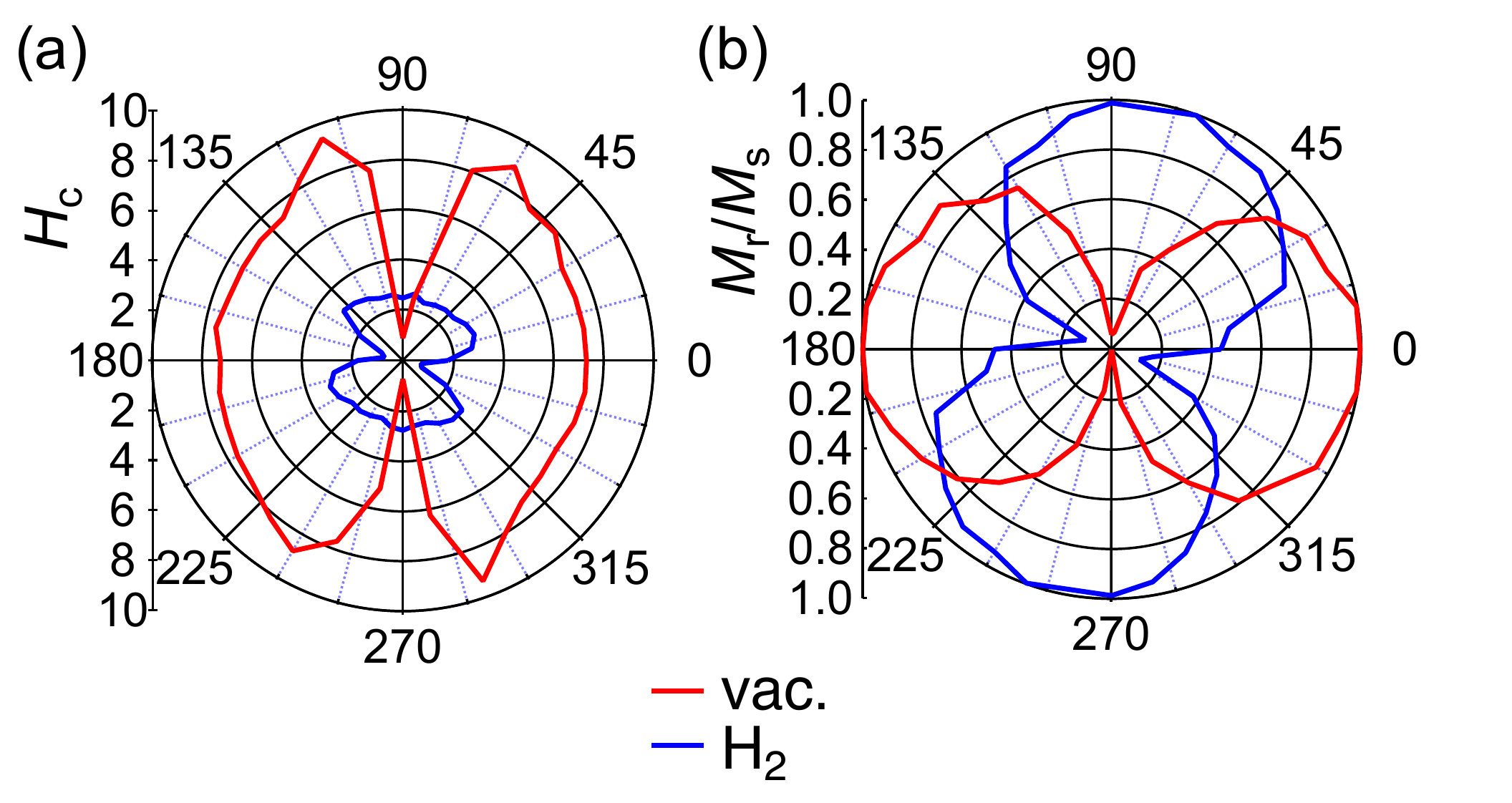}
\caption{The angular dependence of (a) $H_\mathrm{c}$ and (b) $M_\mathrm{r}$/$M_\mathrm{s}$ in vacuum and at $n=32$ after hydrogen exposure, respectively. }
\label{angular}
\end{figure}

The angular dependence of $H_\mathrm{c}$ and $M_\mathrm{r}$/$M_\mathrm{s}$ is displayed in Fig.~\ref{angular}(a) and \ref{angular}(b), respectively. The red curve represents the data in vacuum, whereas the blue curve represents the data at $n=32$ after hydrogen exposure. Angle $0^\circ$ stands for the orientation of the original easy axis. From the data in the figure, the easy axis is rotated to $\sim$75$^\circ$ after hydrogen saturation, as $H_\mathrm{c}$ and $M_\mathrm{r}$/$M_\mathrm{s}$ both reach their maximum at $\sim$75$^\circ$. This new magnetic axis supports a new domain pattern with zigzag wall lines as observed in Fig.~4 at $n=7$ and 32. 

\section{X-ray Crystallography}

\begin{figure}
\centering
\includegraphics[width=0.45\textwidth]{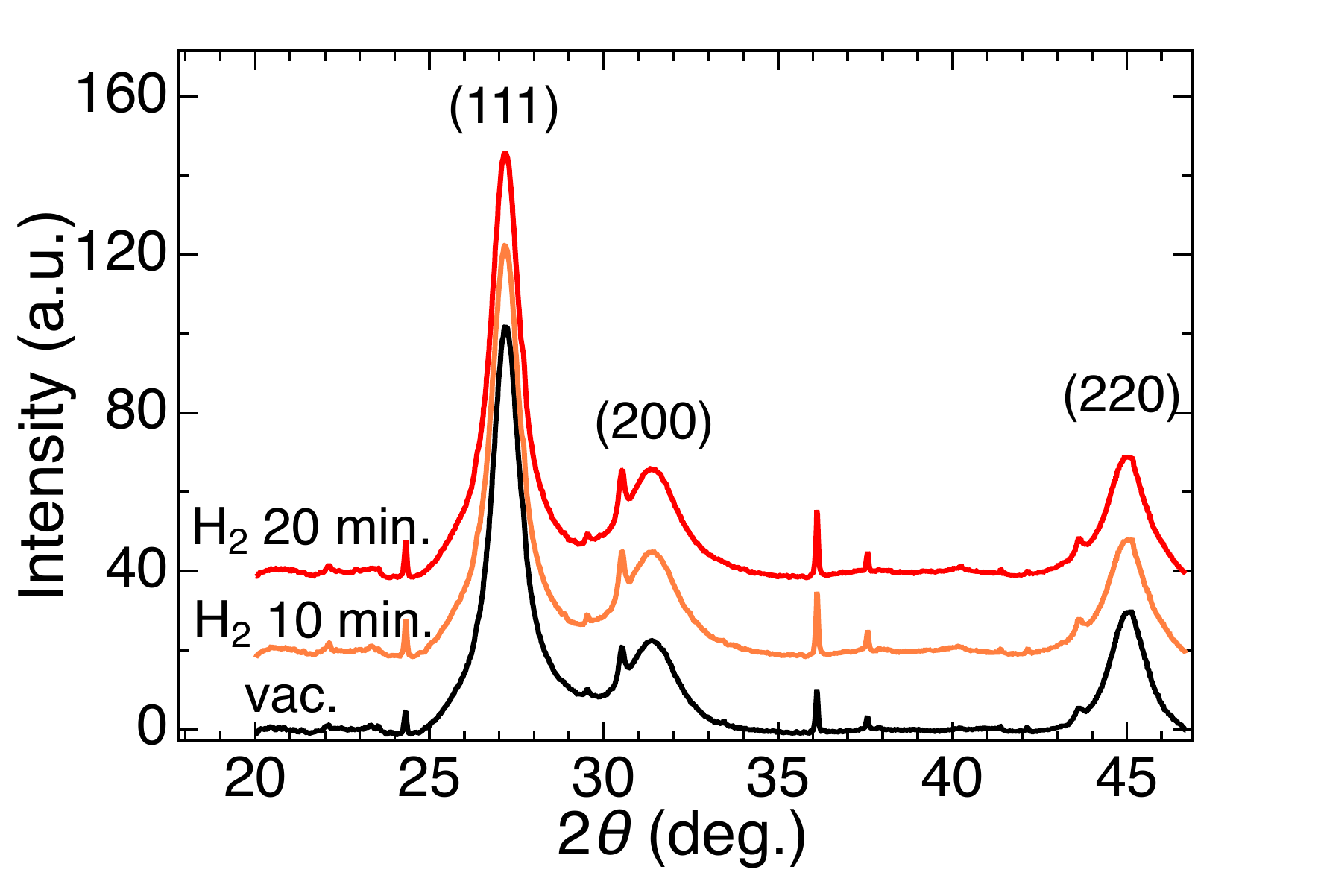}
\caption{XRD of the sample when in vacuum, after 10-min exposure, and after 20-min exposure to hydrogen, respectively.}
\label{XRD}
\end{figure}

\textit{In situ} grazing incident XRD (12 keV) has been performed in the National Synchrotron Radiation Research Center (NSRRC) of Taiwan to examine the crystalline structure of the film before and after exposure to hydrogen, and the results are shown in Fig.~\ref{XRD}. The three curves are the XRD patterns of the FePd film in vacuum, after 10-min exposure, and after 20-min exposure to hydrogen, respectively, and each curve measurement corresponds to 5-min exposure to X-ray. The diffraction patterns are recorded by a MAR345 imaging plate. A 2D diffraction image (not shown) reveals that the FePd thin film is more or less a texture structure. All three measurements are conducted with azimuthal angle ($2\theta$) around 40$^\circ$. The fcc FePd peaks of (111), (200), and (220) are observed in all three curves, and do not change in shape, intensity, or azimuthal angle ($2\theta$) after exposure to hydrogen. This indicates that there is no measurable expansion of the FePd lattice after hydrogen absorption.